\documentclass[english,11pt,aps,prd,a4paper,preprintnumbers,nofootinbib,superscriptaddress,notitlepage]{revtex4}
\usepackage{graphicx}
\usepackage{amsmath}
\usepackage{amssymb}
\usepackage{hyperref}
\usepackage{xcolor}
\usepackage{ulem}
\usepackage{multirow}
\usepackage{comment}
\usepackage{dsfont}
\usepackage{bbold}
\newcommand {\ignore}[1]{}

 \newcommand{\AddrAHEP}{AHEP Group, Institut de F\'{i}sica Corpuscular --C.S.I.C./Universitat de Val\`{e}ncia, Parc Cientific de Paterna.\\C/Catedr\'atico Jos\'e Beltr\'an, 2 E-46980 Paterna (Val\`{e}ncia) - SPAIN}
\newcommand{\AddrCinvestav}{Departamento de F\'{\i}sica, Centro de
  Investigaci{\'o}n y de Estudios Avanzados del IPN\\ Apdo. Postal 14-740 07000 Ciudad de M\'exico, M\'exico}
\newcommand{\AddrTEC}{Laboratorio Interdisciplinario de Investigación Tecnológica-LIIT, Tecnol\'ogico Nacional de M\'exico/ITS de Jerez,
  C.P. 99863 Zacatecas, M\'exico.}
\newcommand{\ident}{\mathbb{1}}

\definecolor{darkred}{rgb}{0.6,0,0}
\definecolor{darkcyan}{RGB}{0, 111, 111}
\definecolor{dgreen}{rgb}{0,0.5,0}

\usepackage{stackengine}

\begin{document}

\title{Searching for generalized neutrino interactions in direct detection experiments with E$\nu$ES}

\author{Jesús Miguel Celestino-Ramírez}\email{jesus.celestino@cinvestav.mx }\affiliation{\AddrCinvestav}
\author{F.~J. Escrihuela}\email{franesfe@alumni.uv.es}\affiliation{\AddrAHEP}
\author{L.~J. Flores}\email{ljflores@jerez.tecnm.mx}\affiliation{\AddrTEC}
\author{O.~G. Miranda}\email{omar.miranda@cinvestav.mx}\affiliation{\AddrCinvestav}
\author{R.~S\'anchez-V\'elez}\email{rsanchezve@ipn.mx}\affiliation{\AddrCinvestav}

\begin{abstract}
{\noindent    We investigate the sensitivity of present and future direct detection experiments to generalized neutrino interactions (GNI) with electrons through elastic neutrino–electron scattering. Using data from LUX-ZEPLIN, PandaX-4T, and XENONnT, we derive constraints on vector, axial-vector, scalar, and tensor effective couplings, and compare them with existing limits. Our results show that current xenon-based detectors already provide competitive bounds, with XENONnT offering the most stringent constraints due to its larger exposure and reduced backgrounds. Among the GNI couplings, the scalar contributions remain more weakly constrained, while tensor interactions yield the strongest limits. We also present projected sensitivities for the DARWIN experiment, showing potential improvements.  These results demonstrate the capability of direct detection experiments, originally designed for dark matter searches, to provide complementary and competitive constraints on generalized neutrino interactions.}
\end{abstract}

\maketitle
\section{Introduction}
The neutrino oscillation discovery motivated an even more exhaustive search for a mechanism that theoretically explains the mass hierarchy pattern, which spans over so many orders of magnitude, particularly, the case of neutrino masses which cluster together at the sub-eV scale. The efforts to theoretically explain the neutrino mass pattern range from the simple minimal extension of the Standard Model (SM) to very elaborated scenarios that may extend beyond the more explored seesaw mechanism. It is common to invoke, for this purpose, several ingredients in addition to the Standard Model, such as new heavy neutral leptons, new gauge mediators, or additional scalars. The phenomenology arising from these models is rich and involves modifications to the Standard Model Lagrangian. A suitable phenomenological framework to study a set of new parameters accounting for physics beyond the SM is the Generalized Neutrino Interactions (GNI) formalism, which introduces effective four-fermion operators in a very general effective Lagrangian. In some sense, the GNI is a generalization of the Non-Standard Interaction (NSI) scheme~\cite{Ohlsson:2012kf,Miranda:2015dra,Farzan:2017xzy,Proceedings:2019qno}. The GNI formalism parametrizes the  additional Beyond the Standard Model (BSM) interactions in terms of 
vector, axial, tensor, scalar, or pseudoscalar contributions,  allowing for the probing of a wide variety of ultraviolet completions, such as extended gauge symmetries, 
  new light gauge bosons, leptoquarks, or vector and scalar mediators~\cite{Schechter1980gr,Mohapatra1986bd,foot:1988aq,hirsch:2004he,Malinsky:2005bi,Grimus:2006nb,Buchmuller:1986zs,Crivellin:2019dwb,Gargalionis:2020xvt}.

Signals of new physics may appear when neutrinos interact with hadrons or charged leptons. GNI could alter the corresponding cross-section and measured energy spectra; their implications have been studied in several experimental searches performed across a wide range of energies.
At low energies, solar neutrino experiments, such as Borexino, and reactor experiments, like TEXONO, have provided constraints through Elastic Neutrino–Electron Scattering (E$\nu$ES)~\cite{Khan:2019jvr,Chen:2021uuw,Coloma:2022umy,Denton:2024upc}. 
Moreover, E$\nu$ES might be a tool for distinguishing between the Dirac or Majorana nature of neutrinos in the presence of GNI~\cite{Rodejohann:2017vup, Marquez:2024tjl}.
Besides, Coherent Elastic Neutrino–Nucleus Scattering (CE$\nu$NS) measurements by COHERENT, PandaX-4T, XENONnT, and CONUS+ have probed GNI using pion decay-at-rest, solar, and reactor fluxes~\cite{Lindner:2016wff,AristizabalSierra:2018eqm,Flores:2021kzl,DeRomeri:2022twg,AristizabalSierra:2024nwf,Liao:2025hcs,Alpizar-Venegas:2025wor,Chattaraj:2025fvx,DeRomeri:2025csu,TEXONO:2025sub}. On the other hand, Neutrino Deep Inelastic Scattering ($\nu$DIS), meson decays, $\beta$ decay, and collider experiments provide limits at the high-energy region~\cite{Han:2020pff,Escrihuela:2021mud, Escrihuela:2023sfb, KATRIN:2024odq}. This broad experimental program highlights the complementarity of different energy scales in testing generalized neutrino interactions.

Direct detection experiments provide a natural extension to the search for generalized neutrino interactions. Originally designed to detect dark matter, their large target masses, ultra-low backgrounds, and excellent sensitivity to electron and nuclear recoil events allow them to probe neutrino interactions at an unprecedented level~\cite{Link:2019pbm, AtzoriCorona:2022jeb,Giunti:2023yha, Maity:2024aji}. 
Using data from LUX-ZEPLIN~\cite{LZ:2019sgr,LZ:2023poo, LZ:2022lsv}, PandaX-4T~\cite{PandaX-4T:2021bab,PandaX:2022ood}, and XENONnT~\cite{XENON:2022ltv}, together with the projected sensitivity of DARWIN~\cite{DARWIN:2020bnc} and a future joint collaboration within the XLZD framework~\cite{Baudis:2024jnk, XLZD:2024nsu}, these detectors complement solar and CE$\nu$NS experiments by testing GNI with extraordinary precision in the sub-MeV regime.

In this work, we examine the sensitivity of current and future direct detection experiments to generalized neutrino  interactions with electrons  \footnote{For the corresponding constraints for quarks from direct detection experiments, the reader can refer to Ref.~\cite{AristizabalSierra:2024nwf}}. Using data from LUX-ZEPLIN, PandaX-4T and XENONnT on electron recoils, we present constraints on different GNI strength parameters, and compared them with the current strongest bounds. We also investigate the potential sensitivity of the future DARWIN experiment. 

The rest of the article is organized as follows. In section \ref{sec:formalism} we introduce the formalism of general neutrino interactions, including the expression for the E$\nu$ES differential cross section with all the possible new contributions, and highlight the relevant terms for direct detection experiments. In section \ref{sec:experimental} we provide a detailed description of the experimental setup, including neutrino flux, detector efficiency, energy resolution and background, as well as a comprehensive explanation of the statistical analysis used. In section \ref{sec:results} we give our results on the constraints for the different GNI parameters. Finally, we present our conclusions in section \ref{sec:conclusions}.
	
\section{GNI formalism}
\label{sec:formalism}
In this section, we introduce the general framework that we will use to phenomenologically model the possible signatures for new physics. We start with the effective Lagrangian describing neutral-current generalized neutrino interactions with electrons, that can be written as~\cite{Bischer:2018zcz,Bischer:2019ttk,Han:2020pff}
\begin{equation}
	\mathcal{L}^\mathrm{NC}_\mathrm{eff}=-\frac{G_{F}}{\sqrt{2}}\sum_{\alpha,\beta}\sum_{j}\overset{(\sim)}{\epsilon^{e,j}_{\alpha\beta}}(\bar{\nu}_{\alpha}\mathcal{O}_{j}\nu_{\beta})(\bar{e}\mathcal{O}^{'}_{j}e)\,,\\
	\label{equation_1}
\end{equation}
where $G_{F}$ is the Fermi constant,  $\nu_{\alpha,\beta}$ is a neutrino with flavor $\alpha,\beta$,  and the couplings $\epsilon^{e,j}_{\alpha\beta}$ provide information about the strength of the new interactions. The operators that account for the different Lorentz structures of the neutrino current are denoted by $\mathcal{O}_{j}$, while in the electron current case we use $\mathcal{O}^{'}_{j}$. The index $j = L, R, S, P, T$, corresponds to left-handed, right-handed, scalar, pseudoscalar, and tensor contributions, respectively. The explicit forms of these operators are given  in Table~\ref {tab:operators}.

The $\tilde{\epsilon}^{e,j}$ parameters require the presence of an incoming flux of right-handed neutrinos. If it were present, it would be suppressed by the smallness of neutrino masses and the left-handedness of weak interactions. Therefore, for the present study, we restrict the analysis to the subset of operators proportional to the $\epsilon^{e,j}$ couplings.

\begin{table}[!hbt]
	\centering
	\begin{tabular}{|ccccc|}
		\hline  \hline
		$\overset{(\sim)}{\epsilon^j}$ & & $\mathcal{O}_{j}$ & &$\mathcal{O}^{'}_{j}$\\
		\hline \hline
		$\epsilon^{L}$ & &$\gamma_{\mu}(\ident-\gamma^{5})$ & &$\gamma^{\mu}(\ident-\gamma^{5})$\\
		\hline
		$\tilde{\epsilon}^{L}$ & &$\gamma_{\mu}(\ident+\gamma^{5})$ & &$\gamma^{\mu}(\ident-\gamma^{5})$\\
		\hline
		$\epsilon^{R}$ & &$\gamma_{\mu}(\ident-\gamma^{5})$ & &$\gamma^{\mu}(\ident+\gamma^{5})$\\
		\hline
		$\tilde{\epsilon}^{R}$ & &$\gamma_{\mu}(\ident+\gamma^{5})$ & &$\gamma^{\mu}(\ident+\gamma^{5})$\\
		\hline
		$\epsilon^{S}$ & &$(\ident-\gamma^{5})$ & & $\ident$\\
		\hline
		$\tilde{\epsilon}^{S}$ & &$(\ident+\gamma^{5})$ & & $\ident$\\
		\hline
		$-\epsilon^{P}$ & &$(\ident-\gamma^{5})$ & &$\gamma^{5}$\\
		\hline
		$-\tilde{\epsilon}^{P}$ & &$(\ident+\gamma^{5})$ & &$\gamma^{5}$\\
		\hline
		$\epsilon^{T}$ & &$\sigma_{\mu\nu}(\ident-\gamma^{5})$ & &$\sigma^{\mu\nu}(\ident-\gamma^{5})$\\
		\hline 
		$\tilde{\epsilon}^{T}$ & &$\sigma_{\mu\nu}(\ident+\gamma^{5})$ & &$\sigma^{\mu\nu}(\ident+\gamma^{5})$\\
		\hline\hline
	\end{tabular}
	\caption{Operators and effective couplings appearing in the Lagrangian described in Eq.~\eqref{equation_1}. All flavor labels have been omitted for simplicity.}
	\label{tab:operators}
\end{table}

From the Lagrangian in Eq.~\eqref{equation_1} and including the SM interactions, taking into account the considerations from above, the cross section for
$\nu_{\alpha}+e^{-}\rightarrow \nu_{\beta}+e^{-}$ takes the form
\begin{equation}
\frac{d\sigma_{\nu_{\alpha,\beta}}}{dT_e}=Z^\mathrm{Xe}_\mathrm{eff}(T_e)\frac{G^{2}_{F}m_{e}}{\pi}\left[A+2B\left(1-\frac{T_e}{E_\nu}\right)+C\left(1-\frac{T_e}{E_\nu}\right)^{2}+D\frac{m_{e}T_e}{E_\nu^{2}}\right],
	\label{eq:crossSec}
\end{equation}
where $m_e$ is the electron mass, $T_e$ denotes the electron recoil kinetic energy,  $E_\nu$ represents the energy of the incoming neutrino and $Z^\mathrm{Xe}_\mathrm{eff}(T_e)$ quantifies the effective  number of electrons which can be ionized at the kinetic energy $T_e$. As for $Z^\mathrm{Xe}_\mathrm{eff}(T_e)$, we use the values given by the authors in~\cite{AtzoriCorona:2022jeb}.  Additionally, the coefficients that encapsulate the contributions from the GNI and the SM are given by
\begin{equation}\label{eq:_coefficients}
\begin{split}
A&=2|(g^{L,\nu_\alpha} \delta_{\alpha\beta} +\epsilon^{e,L}_{\alpha\beta})|^{2}+\frac{1}{4}(|\epsilon^{e,S}_{\alpha\beta}|^{2}+|\epsilon^{e,P}_{\alpha\beta}|^{2})+8|\epsilon^{e,T}_{\alpha\beta}|^{2}-2\Re\left[(\epsilon^{e,S}+\epsilon^{e,P})_{\alpha\beta}\epsilon^{e,T*}_{\alpha\beta}\right],\\
B&=-\frac{1}{4}(|\epsilon^{e,S}_{\alpha\beta}|^{2}+|\epsilon^{e,P}_{\alpha\beta}|^{2})+8|\epsilon^{e,T}_{\alpha\beta}|^{2},\\
C&=2|(g^R \delta_{\alpha\beta} +\epsilon^{e,R}_{\alpha\beta})|^{2}+\frac{1}{4}(|\epsilon^{e,S}_{\alpha\beta}|^{2}+|\epsilon^{e,P}_{\alpha\beta}|^{2})+8|\epsilon^{e,T}_{\alpha\beta}|^{2}+2\Re\left[(\epsilon^{e,S}+\epsilon^{e,P})_{\alpha\beta}\epsilon^{e,T*}_{\alpha\beta}\right],\\
D&=-2\Re\left[(g^{L,\nu_\alpha}\delta_{\alpha\beta} +\epsilon^{e,L}_{\alpha\beta})(g^R \delta_{\alpha\beta} +\epsilon^{e,R*}_{\alpha\beta})\right]+\frac{1}{2}|\epsilon^{e,S}_{\alpha\beta}|^{2}-8|\epsilon^{e,T}_{\alpha\beta}|^{2}.
\end{split}
\end{equation}
The SM contribution is included in coefficients $A$, $C$, and $D$, through the left and right couplings  $g^{L,\nu_{\mu}} = g^{L,\nu_{\tau}} = -1/2 + \sin^2\theta_{W}$, $g^{L,\nu_e} = 1/2 + \sin^2\theta_{W}$ and $g^R = \sin^2\theta_{W}$, which can be written in terms of the vector and axial couplings as $g^L = (g^V + g^A)/2$, and $g^R = (g^V - g^A)/2$. Notice that only the flavor-diagonal left and right GNI couplings produce an interference term with the SM.

From Eq.~\eqref{eq:crossSec}, we can see that the term with the coefficient $D$ is proportional to the ratio $m_e/E_\nu$. This term is suppressed for neutrino energies above a few MeV (see the discussion in~\cite{Escrihuela:2021mud}). However, in direct detection experiments, the main contribution to the solar neutrino background comes from the $pp$ chain, which covers energies up to $\sim 400$ keV, leading to $m_e / E_{\nu} \gtrsim 1 $. Given that, for the purpose of this study, there are no suppressed terms in Eq.~\eqref{eq:crossSec}. Let us point our attention to the scalar and pseudoscalar couplings in Eq.~\eqref{eq:_coefficients}. The $D$ coefficient is the only one where the pseudoscalar coupling does not appear, and since this is the leading term in the cross section, pseudoscalar interactions are negligible in comparison to scalar and tensor contributions. Therefore, we will not include the pseudoscalar interaction in the following analysis.

In order to keep the analysis simpler, we will be considering all GNI parameters as real, i.e. all CP-violating phases are set to zero,  therefore we will have 24 free parameters: six possible neutrino flavor combinations, and four different Lorentz invariants $V,A,S$, and $T$.

\section{Experimental description and analysis}
\label{sec:experimental} 
In order to perform our analysis, we consider data from dark matter direct detection experiments, such as PandaX-4T~\cite{PandaX-4T:2021bab, PandaX:2022ood}, LUX-ZEPLIN~\cite{LZ:2019sgr, LZ:2023poo, LZ:2022lsv}, and XENONnT~\cite{XENON:2010xwm, XENON:2022ltv}. Although these experiments are primarily designed to search for dark matter, they also provide valuable opportunities to study neutrino properties by analyzing the $\mathrm{E\nu ES}$ processes. In particular,
\begin{itemize}
\item LUX-ZEPLIN is located 4850 ft underground in the Davis Cavern at the Sanford Underground Research Facility (SURF) in Lead, South Dakota, USA~\cite{Mount:2017qzi}. This experiment employs a detector built with 10 tonnes of liquid xenon, where a background of ${^8}\text{B}$ solar neutrinos is expected.
\item PandaX-4T is a detector with a 3.7-tonne liquid as target and located  in the China Jinping Underground Laboratory (CJPL)~\cite{Kang:2010zza, Zhao:2020ezy}. In this setup, ${^8}\text{B}$ solar neutrinos can induce coherent neutrino–nucleus scattering with xenon nuclei. The corresponding background is estimated to be $0.6\pm 0.3$ events~\cite{Ruppin:2014bra}.
\item XENONnT is a dark matter experiment with a core of 8.5 tonnes of liquid xenon. It is placed on the INFN Laboratori Nazionali del Gran Sasso (LNGS) in Italy~\cite{GranSasso_web}. 
\end{itemize}

As it has been mentioned previously, the ultimate goal of these experiments is the direct detection of dark matter. Achieving this requires a precise characterization of all relevant backgrounds. Among the most significant of these backgrounds are neutrino interactions, which must be carefully accounted for as follows: 
\begin{equation}
 N^{L}_k = N^\mathrm{E \nu ES}_k + B_k,
\label{eq:numb_ev}    
\end{equation}
where the super index $L$ refers to the experiment, the k index refers to the energy bin, and the second term $B_k$ is the background contribution. The $N^\mathrm{E\nu ES}_k$ is the contribution from solar neutrinos and we can obtain it from 
\begin{equation}
N^\mathrm{E\nu ES}_k = N \int^{T^{k+1}_e}_{T^{k}_e} dT_e\; A(T_e) \int^{\infty}_0 dT^{\prime}_{e}\; R(T_e,T_e^\prime) \sum_{i=pp,^7\text{Be}} \int^{E^{\text{max}}_{\nu, i}}_{E^{\text{min}}_{\nu}}dE_{\nu} \sum_{{\alpha,\beta=e,\mu,\tau}} \Phi^{i}_{\nu_\alpha} (E_{\nu})\frac{d\sigma_{\nu_{\alpha,\beta} } }{dT^{\prime}_e} \, .
\label{eq:EvES}
\end{equation}
In this equation, $T_e$ and $T^{\prime}_e$ represent the reconstructed and true electron recoil energies, respectively. Further, the electron-neutrino cross section per xenon atom is given by 
$\frac{d\sigma_{\nu_{\alpha,\beta} } }{dT^{\prime}_e}$, and the solar neutrino fluxes  $\Phi^{\prime i}_{\nu_\alpha}$ are expressed as
\begin{equation}
\Phi^i_{\nu_e}=\Phi^{\prime i}_{\nu_e}P_{ee},\;\; \Phi^i_{\nu_\mu}=\Phi^{\prime i}_{\nu_e}(1-P_{ee})\cos^2\theta_{23},\:\; \Phi^i_{\nu_\tau}=\Phi^{\prime i}_{\nu_e}(1-P_{ee})\sin^2\theta_{23}.
\end{equation}
Here, $\Phi^{\prime i}_{\nu_e}$ are the fluxes of electron neutrinos produced by thermonuclear reactions in the core of the sun~\cite{Bahcall:1987jc,Bahcall:1994cf,Bahcall:1996qv}. The three-neutrino survival probability, $P_{ee}$, for solar neutrinos reaching the detector considering the dominant $pp$ and $^7\text{Be}$ fluxes, is given by~\cite{Chen:2016eab}
\begin{equation}
P_{ee}=\sin^4\theta_{13}+P^{2\nu}\cos^4\theta_{13},
\end{equation}
where $P^{2\nu}=\left(1 - \sin^2(2\theta_{12}) /2 \right) \approx 0.55$ is the survival probability of $\nu_e$  in the two-neutrino oscillation scheme~\cite{Chen:2016eab}.
The values of the mixing angles $\theta_{13}$ and $\theta_{23}$ were taken from ~\cite{ParticleDataGroup:2024cfk}. When evaluating the integral over $E_\nu$ in Eq.~\eqref{eq:EvES}, the corresponding variation of energy goes from $E^\mathrm{min}_{\nu}$ to $E^\mathrm{max}_{\nu}$, where the minimum energy is given by 
\begin{equation}
E^\mathrm{min}_{\nu}= \frac{T^{\prime}_e + \sqrt{2 m_e T^{\prime}_e + {T^{\prime}_e}^2}}{2},
\label{eq:emin}
\end{equation}
while the value of $E^\mathrm{max}_{\nu}$ depends on the production process of the incident neutrinos. Finally, following from right to left, $R(T_e, T^{\prime}_e)$ and $A(T_e)$ are the resolution and efficiency of the detector, which are different for each experiment. $N$ is a normalization constant that includes peculiarities of each experiment, such as exposure time and detector volume. The different values of detector efficiencies of Eq.~(\ref{eq:EvES}) are taken from~\cite{PandaX:2022ood, XENON:2022ltv, LZ:2023poo}, while for energy resolution, we use the same function that we can find in~\cite{AtzoriCorona:2022jeb, A:2022acy} for LUX-ZEPLIN, in~\cite{XENON:2020rca} for XENONnT, and in~\cite{PandaX:2022ood} for PandaX-4T.

In order to carry out our analysis, we will compare our theoretical calculation of the number of events, $N^L_k$, with the data published by the different experimental collaborations, $D^L_k$, through the $\chi^2$ function. In particular, we use data from~\cite{LZ:2022lsv} for LUX-ZEPLIN,~\cite{PandaX:2022ood} for PandaX-4T, and~\cite{XENON:2022ltv} for XENONnT.

Nevertheless, this analysis is not the same for all three experiments. Because some bins contain too few data points, for LUX-ZEPLIN and PandaX-4T we use a Poisson distribution to perform our calculation:
\begin{equation}
\chi^{2}_{L}~=~\min\limits_{\alpha,\beta} \left\{2 \left(\sum_{k} N^{L}_k - D^{L}_{k} + D^{L}_{k} \; \text{log} \frac{D^{L}_{k}}{N^{L}_{k}} \right) + \left(\frac{\alpha}{\sigma_{\alpha}}\right)^2 + \left(\frac{\beta}{\sigma_{\beta}}\right)^2 \right\} \,,
\label{eq:pois_chi}
\end{equation}
where $\alpha$ and $\beta$ are nuisance coefficients for solar neutrino fluxes and background neutrinos respectively, being $\sigma_{\alpha}$ the $\alpha$ uncertainty taken from~\cite{Villante:2020adi} and $\sigma_{\beta}$ the $\beta$ uncertainty taken from~\cite{LZ:2022ysc} for LUX-ZEPLIN, and from~\cite{PandaX:2022ood} for PandaX-4T. 

On the other hand, for the analysis of the XENONnT data, we use the function
\begin{equation}
\chi^{2}_{X}~=~\min\limits_{\alpha,\beta} \left\{ \sum_{k} \left(\frac{N^\mathrm{XENONnT}_k - D^\mathrm{XENONnT}_{k}}{\sigma_{k}}\right)^2 + \left(\frac{\alpha}{\sigma_{\alpha}}\right)^2 + \left(\frac{\beta}{\sigma_{\beta}}\right)^2 \right\} \,,
\label{eq:norm_chi}
\end{equation}
in which we find the same components as in Eq.~(\ref{eq:pois_chi}) for LUX-ZEPLIN and PandaX-4T. In this case, values have been taken from~\cite{XENON:2022ltv}.

For our analysis, we initially considered each experiment separately. After that, a combined analysis was also performed, taking into account the data from each experiment. Because all detectors observe neutrinos from the same source, the solar flux uncertainty is correlated among all three datasets; we have verified the impact of this correlation on the limits and found that it is negligible. Our results are summarized in Table~\ref{tab:table_1}.

Looking ahead, we analyze the sensitivity of the DARWIN~\cite{DARWIN:2020bnc} future experiment to GNI parameters. In this case, the event rate calculation at DARWIN follows the same formalism as employed in Eqs.~(\ref{eq:numb_ev}) and (\ref{eq:EvES}).

The background components are modeled using the spectra from~\cite{DARWIN:2020bnc}, normalized to the relevant exposure. For this calculation, we adopt the same resolution function and detector efficiency as XENONnT, assuming constant efficiency for $T_e > T_{e,max}^{\text{XENONnT}}$. Under these assumptions, we reproduce the $\mathrm{E\nu ES}$ spectra, confirming the validity of our approach. It should be mentioned that a DARWIN sensitivity corresponding to an exposure of 30 ty has been taken into account.

\section{Results}
\label{sec:results}

In this section we present a synthesis of the results derived from the ${\chi}^2$ statistical analysis performed on the experimental dataset discussed previously, under the hypothesis of generalized neutrino interactions, considering one parameter at a time.
We will show constraints on vector, axial, scalar, and tensor couplings considering different experiments. 
 First, we analyzed each experimental dataset independently. Afterwards, we performed a comprehensive fit combining different datasets to obtain bounds on the relevant GNI coupling parameters.  Each experiment is characterized by sensitivity to six independent observables (three diagonal and three non-diagonal) associated with vector, axial, scalar, and tensor interactions. In this analysis, we assume that only one observable may be different  from zero while the others vanish, thereby isolating individual contributions. \\

\begin{figure}
	\centering
	\includegraphics[width=0.3\textwidth]{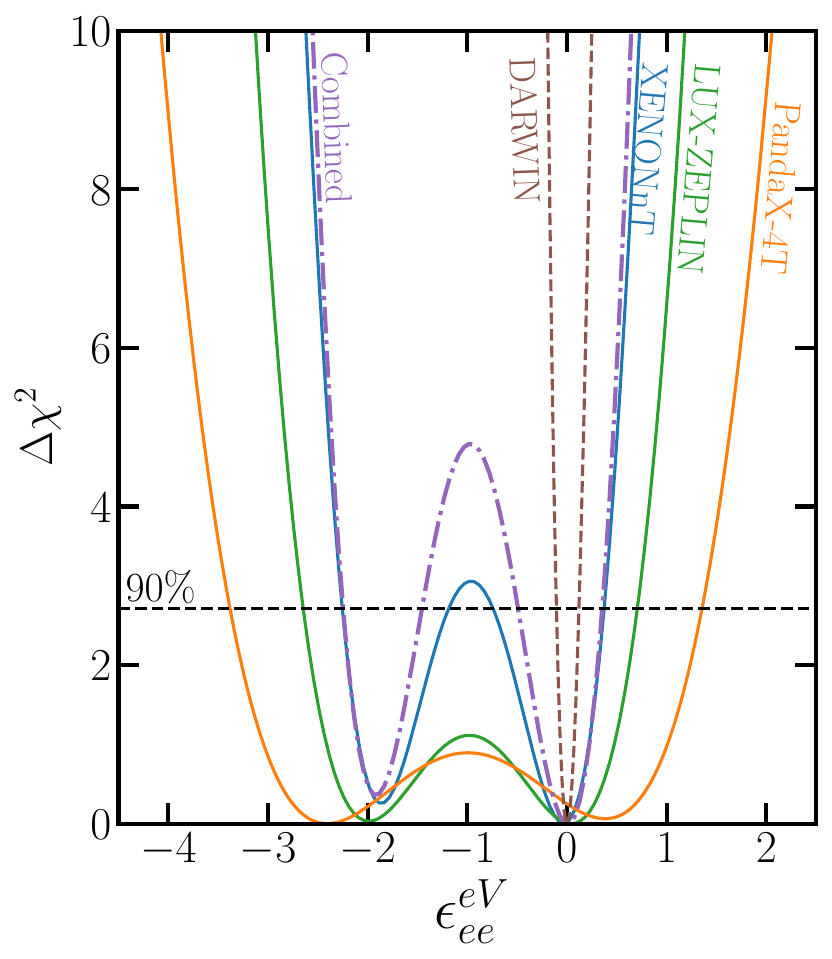}
	\includegraphics[width=0.295\textwidth]{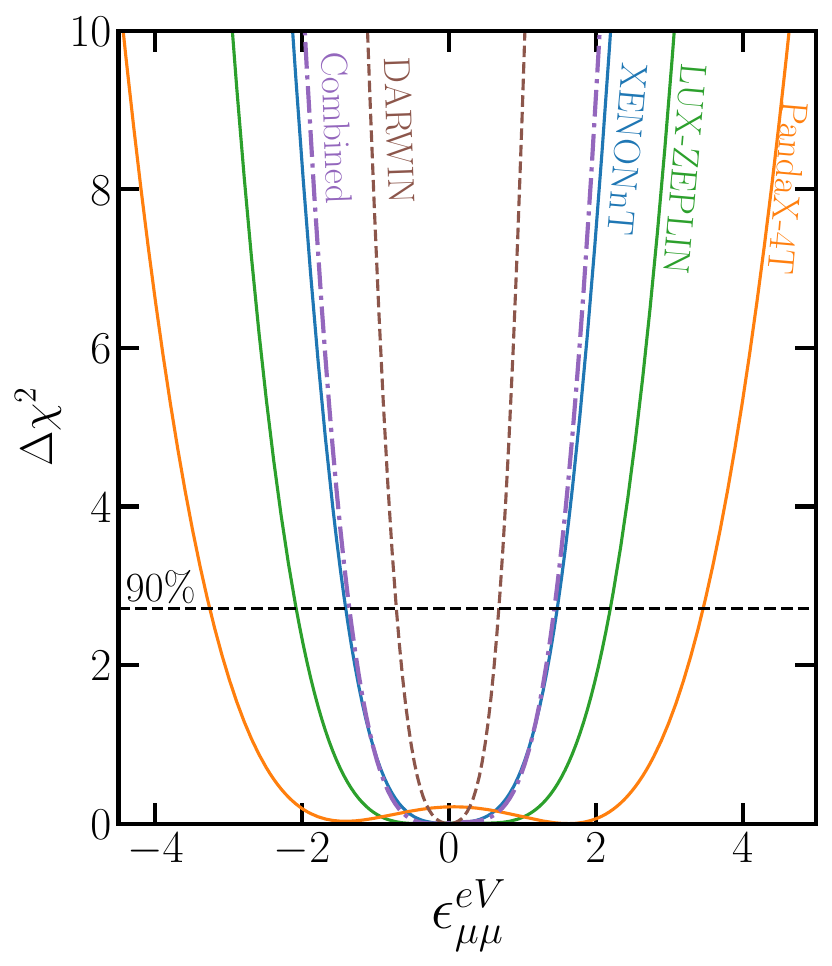}
	\includegraphics[width=0.3\textwidth]{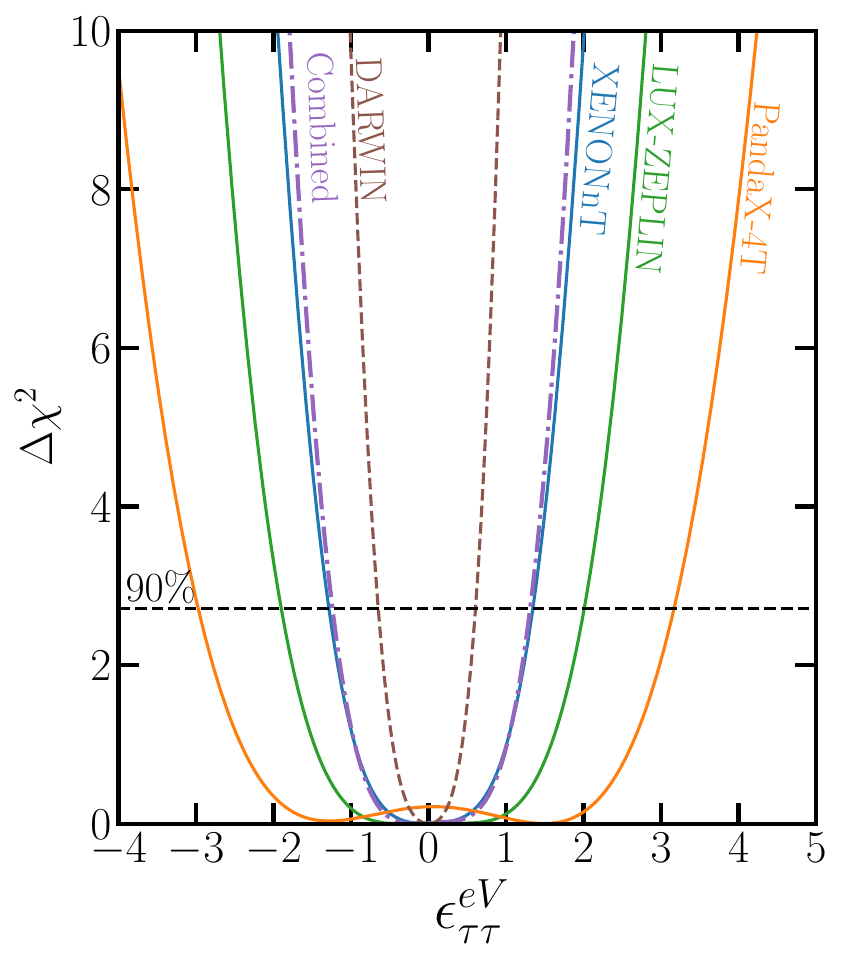} \\
	\includegraphics[width=0.3\textwidth]{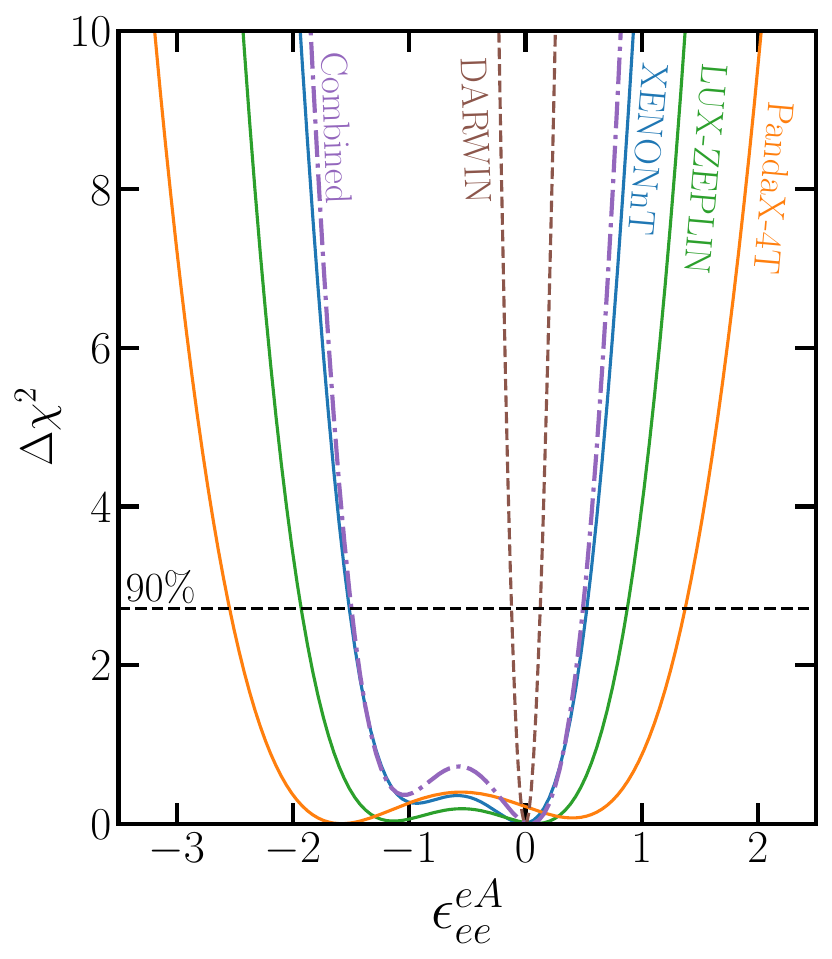}
	\includegraphics[width=0.3\textwidth]{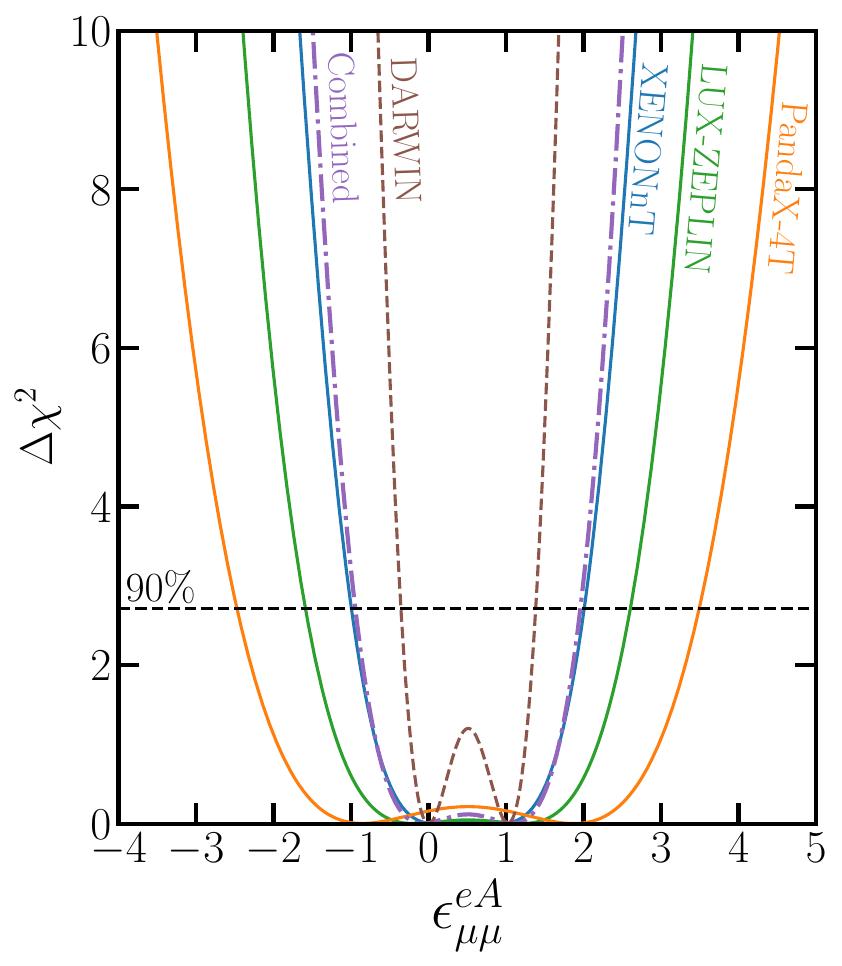}
	\includegraphics[width=0.3\textwidth]{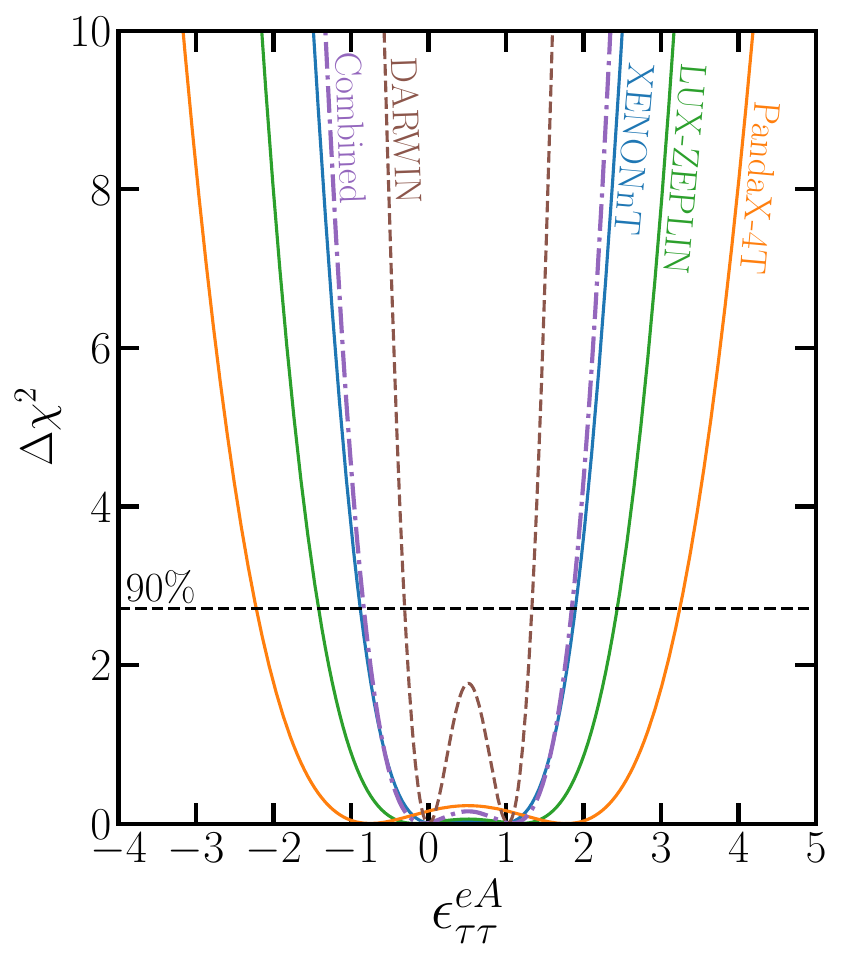}
	\caption{Vector and axial-vector GNI parameters $\Delta\chi^2$ profiles obtained from our analyses. The upper (lower) row corresponds to the vector (axial-vector) flavor-conserving parameters. The solid lines stand for the results using data from PandaX-4T, LUX-ZEPLIN, and XENONnT. The combined analysis for these three experiments is shown with the dashed-dotted line. We also present the allowed region for the coming experiment DARWIN with a dashed line.}
\label{fig:v-a}
\end{figure}

Using data from the previously described dark matter search experiments we perform a goodness of fit analysis as described in the previous section. We present, in Fig.~\ref{fig:v-a}, the $\Delta\chi^2$ profiles for both vector and axial-vector GNI couplings. These plots focus exclusively on the diagonal couplings $\epsilon^{e,X}_{\alpha\alpha}$, where $\alpha = e, \mu, \tau$. 

As detailed in Eq. (3), interference terms with the SM contributions occur in diagonal interactions involving the vector and axial-vector GNI parameters. These interference terms may result in cancellations among the contributions, and the corresponding regions are generally larger than those involving only non-diagonal interactions. As observed, the vector GNI restrictions we found in this work are less stringent than those reported by the authors in~\cite{Coloma:2023ixt}, who included a global analysis with neutrino oscillation data.  As for the axial-vector GNI coupling, the bounds found in the combined analysis are rather close, and then competitive, with the bounds in~\cite{Coloma:2023ixt}.
 \\

\begin{figure}
	\centering
	\includegraphics[width=0.3\textwidth]{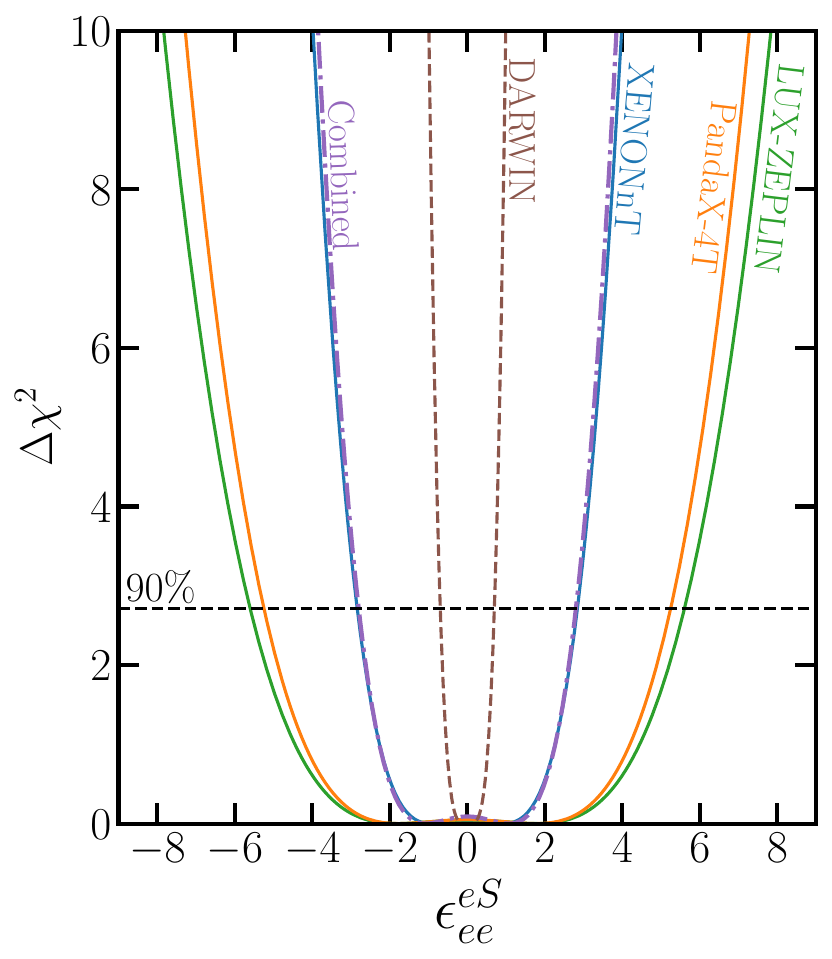}
	\includegraphics[width=0.31\textwidth]{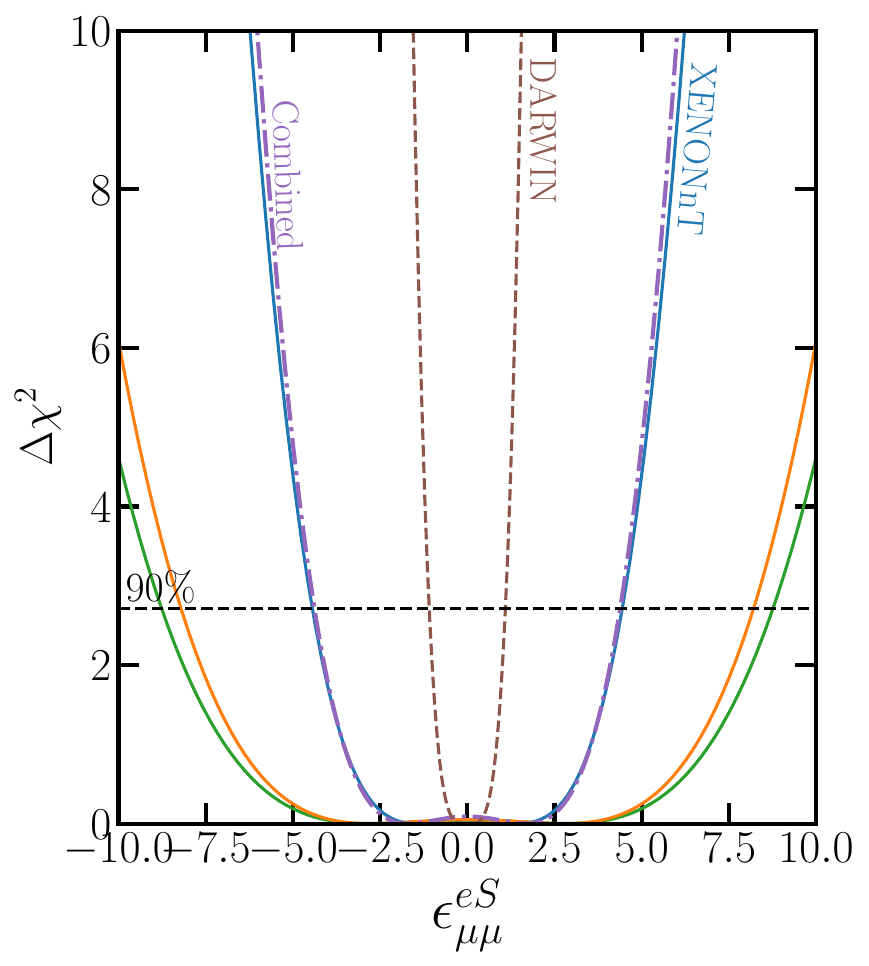}
	\includegraphics[width=0.31\textwidth]{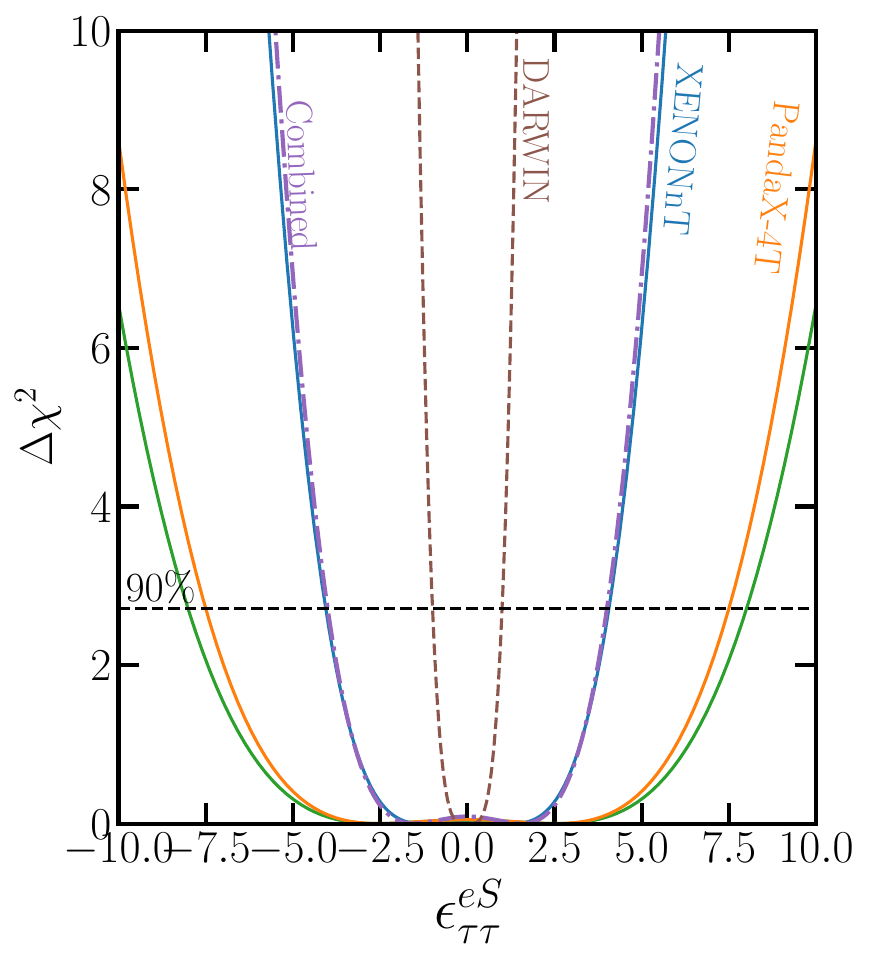} \\
	\includegraphics[width=0.3\textwidth]{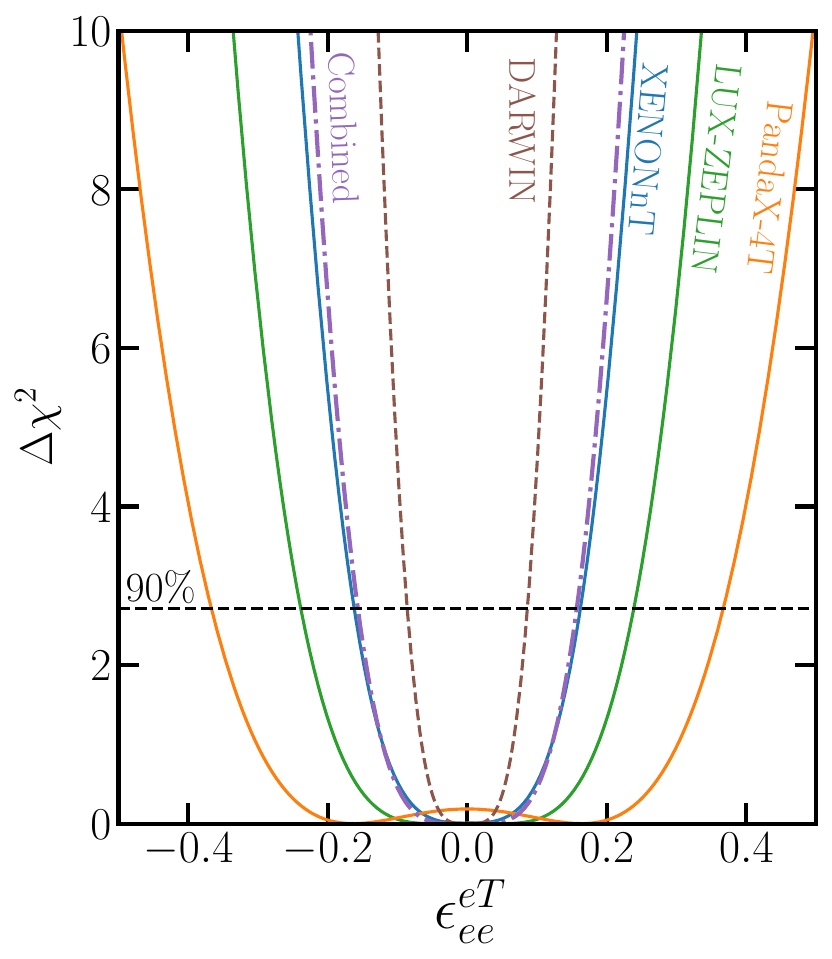}
	\includegraphics[width=0.306\textwidth]{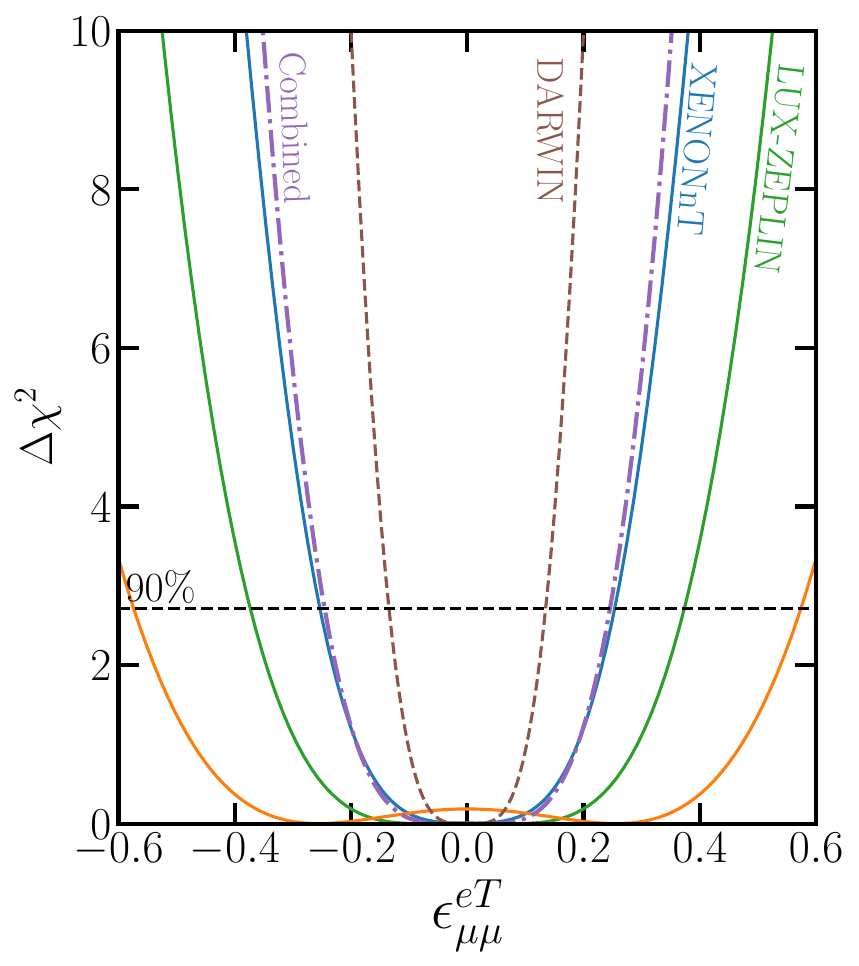}
	\includegraphics[width=0.306\textwidth]{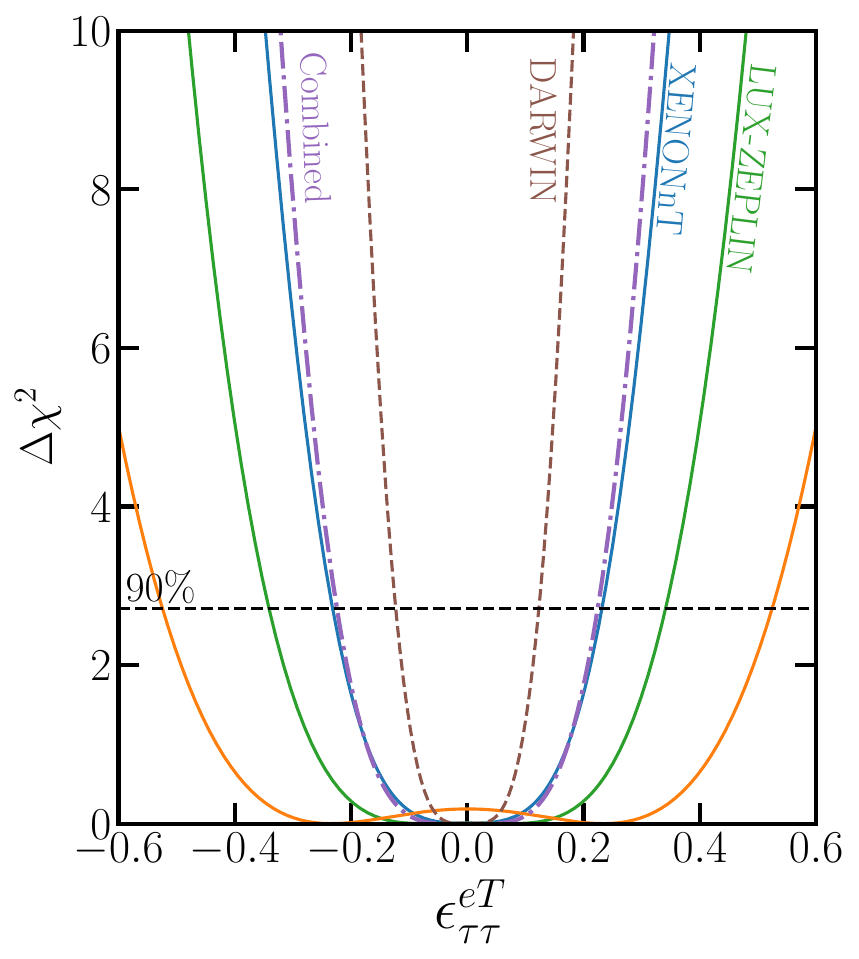}
	\caption{Scalar and tensor GNI parameters $\Delta\chi^2$ profiles obtained from our analyses. The upper (lower) row corresponds to the scalar (tensor) flavor-conserving parameters. The solid lines stand for the results using data from PandaX-4T, LUX-ZEPLIN, and XENONnT. The combined analysis for these three experiments is shown with the dashed-dotted line. We also present the projected sensitivity for the future DARWIN  experiment with a dashed line.}
\label{fig:s-t}
\end{figure}

In Fig.~\ref{fig:s-t}, the $\Delta \chi^2$ profiles for scalar and tensor diagonal GNI parameters are displayed. For the scalar GNI coupling, the LUX-ZEPLIN data provide the weaker constraints, although they are relatively close to the limits set by PandaX-4T. As established in Section~\ref{sec:formalism}, the contributions of the tensor parameter to the cross section are approximately ten times greater, resulting in the most stringent limits for the GNI couplings. As mentioned before, the XENONnT experiment provides the best limits analyzed in this work, due to its large statistics and lower background rates. In contrast, the PandaX-4T dataset has the lowest statistics, reporting 30 events, compared to 52 events from the LUX-ZEPLIN collaboration and around 85 events from XENONnT. On the other hand, it is noteworthy that the neutrino electron GNI couplings $\epsilon^{eX}_{ee}$ exhibit the most restricted upper bounds. This is due to the fact that the processes $\nu_{e}+e^{-}\rightarrow \nu_{e}+e^{-}$ receives contribution from both $W$ and $Z$ bosons, while the process $\nu_{\mu(\tau)}+e^{-}\rightarrow \nu_{\mu(\tau)}+e^{-}$ is mediated only by the $Z$ boson within the SM framework. Additionally, the significant survival probability of electron neutrinos also plays a role in this limitation. It is important to recall that we found different constraints for muon and tau neutrinos, as we do not assume maximal atmospheric mixing i.e. we implement $\sin^2\theta_{23}=0.545$ (normal order). In Table~\ref{tab:table_1}, we report the 90 \% confidence level (CL) upper bounds on the vector, axial-vector, scalar, and tensor parameters for each experiment. Notice that tensor bounds are generally more stringent than scalar ones,  reflecting the approximately tenfold enhancement of tensor contributions within the relevant cross-section expressions. The combined analysis is also presented in Table~\ref{tab:table_1}, demonstrating that the most constraining experimental limits are provided by XENONnT, attributable to its superior statistical power resulting from a larger volume of collected data. Finally, in Table~\ref{tab:table_1}, we present the projected limits derived from the proposed DARWIN experiment assuming an exposure of 30 ty. We observe that the bounds for vector, axial-vector and tensor parameters are improved by a factor of approximately 2 compared with the combined analysis, while the scalar limits are improved by almost a factor of 4. These limits indicate significantly stronger constraints compared to those attainable with current experiments. 

It is worth mentioning that the bounds found in this analysis can be complementary to other studies conducted with data from solar neutrino experiments. Namely, in~\cite{Coloma:2022umy}, the authors  derived limits for both vector and axial vector neutrino couplings with electrons based on the Borexino experiment data. In this case, the constraints are approximately of  $\epsilon^{eA}_{\alpha\alpha}$  and $\epsilon^{eA}_{\alpha\beta}$ $<\mathcal{O}(1)$ order,
  which are comparable to those obtained in this work.

\begin{table}[!hbt]
	\centering
	\begin{tabular}{|c|ccc||c||cc|}
		\hline  \hline
		Parameter                         &PandaX-4T             &LUX-ZEPLIN   	     	&XENONnT          	&Combined			  &Darwin & Current best limit\\
		\hline 
       $\epsilon^V_{ee}$	       &$(-3.38, 1.36)$	   &$(-2.65, 0.71)$  &$(-2.25,-1.19)\cup$  &$(-2.25,-1.46)\cup$   &$(-0.11, 0.12)$ &$(-0.13, 0.10)$ \\
                															& &   &$(-0.74, 0.37)$         &$(-0.49, 0.35)$  & & \\ 
        $\epsilon^V_{\mu\mu}$	  &$(-3.26, 3.46)$   &$(-2.08, 2.20)$  &$(-1.41, 1.47)$		&$(-1.36, 1.44)$ 	       &$(-0.72, 0.68)$ &$(-0.2, 0.10)$\\        				
        $\epsilon^V_{\tau\tau}$	  &$(-2.97, 3.17)$	   &$(-1.90, 2.01)$  &$(-1.29, 1.35)$		&$(-1.24, 1.31)$ 	       &$(-0.65, 0.61)$ &$(-0.17,0.093)$\\												
		$|\epsilon^V_{e\mu}|$	  &$<1.81$		   &$<1.15$		   &$<0.78$		          &$<0.75$		            &$<0.41$ &$(-0.097, 0.011)$\\
		$|\epsilon^V_{e\tau}|$	  &$<1.76$     	   &$<1.12$		   &$<0.75$		          &$<0.73$		            &$<0.40$ &$(-0.18, 0.08)$\\
		$|\epsilon^V_{\mu\tau}|$&$<2.27$		   &$<1.44$		   &$<0.97$		          &$<0.94$		            &$<0.51$ &$(-0.0063, 0.016)$\\
		\hline 
		$\epsilon^A_{ee}$	       &$(-2.55, 1.37)$	   &$(-1.93, 0.88)$  &$(-1.52, 0.53)$	     &$(-1.50, 0.49)$	       &$(-0.13, 0.13)$ &$(-0.13, 0.11)$\\
		$\epsilon^A_{\mu\mu}$	  &$(-2.47, 3.49)$	   &$(-1.59, 2.60)$  &$(-0.99, 2.01)$		&$(-0.94, 1.96)$      	  &$(-0.36, 1.38)$ &$(-0.70, 1.2)$\\
		$\epsilon^A_{\tau\tau}$	 &$(-2.22, 3.24)$	   &$(-1.42, 2.43)$  &$(-0.88, 1.89)$		&$(-0.84, 1.85)$		 &$(-0.31, 1.33)$ &$(-0.53, 1.0)$\\	
		$|\epsilon^A_{e\mu}|$	  &$<1.58$		   &$<1.09$     	   &$<0.76$		          &$<0.73$	          	       &$<0.38$ &$(-0.41, 0.40)$\\
		$|\epsilon^A_{e\tau}|$	  &$<1.54$		   &$<1.07$		   &$<0.74$		          &$<0.71$	           	  &$<0.37$ &$(-0.36, 0.36)$\\
		$|\epsilon^A_{\mu\tau}|$&$<1.98$		   &$<1.37$	        &$<0.95$		          &$<0.92$		            &$<0.47$ &$(-0.79, 0.81)$\\
		\hline 
		$|\epsilon^S_{ee}|$	      &$<5.25$		       &$<5.60$		   &$<2.84$	           	&$<2.80$	                &$<0.70$      &$<0.38$\\
		$|\epsilon^S_{\mu\mu}|$&$<8.21$	   	  &$<8.76$	        &$<4.44$				&$<4.38$ 			 	&$<1.09$            &$<0.31$\\
		$|\epsilon^S_{\tau\tau}|$&$<7.45$		  &$<8.01$		   &$<4.06$				&$<4.00$				&$<1.00$             &$<0.40$\\
		$|\epsilon^S_{e\mu}|$	 &$<4.42$		       &$<4.72$		   &$<2.39$		          &$<2.36$	                &$<0.59$     &$<0.25$\\
		$|\epsilon^S_{e\tau}|$	 &$<4.30$		       &$<4.60$		   &$<2.33$		          &$<2.30$				&$<0.57$          &$<0.28$\\
		$|\epsilon^S_{\mu\tau}|$&$<5.54$		  &$<5.91$		   &$<2.99$			     &$<2.96$				&$<0.74$			&$<0.25$\\
		\hline
		$|\epsilon^T_{ee}|$	      &$<0.367$		  &$<0.239$		   &$<0.162$			     &$<0.157$			&$<0.086$			&$<0.07$\\
		$|\epsilon^T_{\mu\mu}|$&$<0.574$		  &$<0.373$		   &$<0.253$				&$<0.245$ 			&$<0.135$		&$<0.03$	\\
		$|\epsilon^T_{\tau\tau}|$&$<0.524$		  &$<0.341$		   &$<0.232$				&$<0.224$			&$<0.123$			&$<0.12$\\
		$|\epsilon^T_{e\mu}|$	 &$<0.309$		  &$<0.201$		   &$<0.137$			     &$<0.132$			&$<0.073$			&$<0.03$\\
		$|\epsilon^T_{e\tau}|$	 &$<0.301$		  &$<0.196$		   &$<0.133$				&$<0.129$			&$<0.071$			&$<0.07$\\
		$|\epsilon^T_{\mu\tau}|$&$<0.387$		  &$<0.252$		   &$<0.171$				&$<0.166$			&$<0.091$			&$<0.03$\\
		\hline \hline
	\end{tabular}
	\caption{90\% C.L. bounds on the 24  GNI parameters considered in this work. We show the results for  PandaX-4T, Lux-Zeplin, and XENONnT data sets, as well as the limits from a combined analysis. Expected bounds from the upcoming DARWIN experiment are also shown. Current best limits for vector and axial couplings come from a global analysis derived from oscillation data including the effect in the electron scattering~\cite{Coloma:2023ixt}, whereas the limits for scalar and tensor interactions are taken from~\cite{Escrihuela:2021mud}.}
	\label{tab:table_1}
\end{table}

\section{Conclusions}
\label{sec:conclusions}

In this work, we have investigated the sensitivity of present and future dark matter direct detection experiments to generalized neutrino interactions with electrons through elastic neutrino–electron scattering. Using data from LUX-ZEPLIN, PandaX-4T, and XENONnT, we derived constraints on vector, axial-vector, scalar, and tensor GNI couplings. We also provided projections for the DARWIN experiment, which will reach unprecedented exposures.

Our analysis shows that current xenon-based detectors already provide competitive bounds on GNI parameters, complementary to those obtained from solar neutrino experiments such as Borexino, CE$\nu$NS, and reactor experiments, as well as high-energy probes, including neutrino deep inelastic scattering and collider measurements. In particular, we find that
vector and axial-vector couplings are constrained to a similar level to existing global limits, though with some degeneracies arising from interference with Standard Model contributions.
On the other hand, scalar interaction bounds are weaker, but DARWIN is expected to improve these limits by almost a factor of four. Tensor interactions give the most stringent constraints among the new couplings, thanks to their enhanced contribution to the cross section. In particular, current XENONnT data provide the best limits.
Combined analyses of existing datasets strengthen the constraints, with XENONnT dominating thanks to its larger statistics and lower backgrounds.

In addition, DARWIN will significantly improve sensitivity to all GNI couplings, with expected gains of approximately a factor of two for vector, axial-vector, and tensor couplings, and even stronger improvements for scalar contributions. This demonstrates the significant potential of direct detection experiments, originally designed for dark matter searches, to probe neutrino physics beyond the Standard Model.

Our results demonstrate that direct detection experiments constitute a powerful and complementary tool in the global effort to constrain generalized neutrino interactions, further closing the parameter space for new physics scenarios.

\section{Acknowledgments}

This work is supported by the project SECIHTI  CBF-2025-I-1589 and  DGAPA UNAM grant PAPIIT IN111625. The work of O. G. M., L. J. F., and R. S. V. has  been supported by SNII (Sistema Nacional de Investigadoras e Investigadores, Mexico). R. S. V. also thanks SECIHTI for a postdoctoral grant.

\bibliographystyle{utphys} 
\bibliography{merged22}

@article{Link:2019pbm,
	author = "Link, Jonathan M. and Xu, Xun-Jie",
	title = "{Searching for BSM neutrino interactions in dark matter detectors}",
	eprint = "1903.09891",
	archivePrefix = "arXiv",
	primaryClass = "hep-ph",
	doi = "10.1007/JHEP08(2019)004",
	journal = "JHEP",
	volume = "08",
	pages = "004",
	year = "2019"
}

@article{Lindner:2016wff,
	author = "Lindner, Manfred and Rodejohann, Werner and Xu, Xun-Jie",
	title = "{Coherent Neutrino-Nucleus Scattering and new Neutrino Interactions}",
	eprint = "1612.04150",
	archivePrefix = "arXiv",
	primaryClass = "hep-ph",
	doi = "10.1007/JHEP03(2017)097",
	journal = "JHEP",
	volume = "03",
	pages = "097",
	year = "2017"
}

@article{Giunti:2023yha,
	author = "Giunti, Carlo and Ternes, Christoph A.",
	title = "{Testing neutrino electromagnetic properties at current and future dark matter experiments}",
	eprint = "2309.17380",
	archivePrefix = "arXiv",
	primaryClass = "hep-ph",
	doi = "10.1103/PhysRevD.108.095044",
	journal = "Phys. Rev. D",
	volume = "108",
	number = "9",
	pages = "095044",
	year = "2023"
}

@article{Maity:2024aji,
	author = "Maity, Tarak Nath and Boehm, Celine",
	title = "{First constraint on the weak mixing angle using direct detection experiments}",
	eprint = "2409.04385",
	archivePrefix = "arXiv",
	primaryClass = "hep-ph",
	doi = "10.1103/1lgl-bx7v",
	journal = "Phys. Rev. D",
	volume = "112",
	number = "5",
	pages = "053001",
	year = "2025"
}

@article{Chen:2021uuw,
	author = "Chen, Zikang and Li, Tong and Liao, Jiajun",
	title = "{Constraints on general neutrino interactions with exotic fermion from neutrino-electron scattering experiments}",
	eprint = "2102.09784",
	archivePrefix = "arXiv",
	primaryClass = "hep-ph",
	doi = "10.1007/JHEP05(2021)131",
	journal = "JHEP",
	volume = "05",
	pages = "131",
	year = "2021"
}

@article{Marquez:2024tjl,
	author = "M{\'a}rquez, Juan Manuel and Roig, Pablo and Salinas, M{\'o}nica",
	title = "{{\ensuremath{\nu}}e {\textrightarrow} {\ensuremath{\nu}}e scattering with massive Dirac or Majorana neutrinos and general interactions}",
	eprint = "2401.14305",
	archivePrefix = "arXiv",
	primaryClass = "hep-ph",
	doi = "10.1007/JHEP05(2024)227",
	journal = "JHEP",
	volume = "05",
	pages = "227",
	year = "2024"
}

@article{TEXONO:2025sub,
	author = "Karada{\u{g}}, S. and others",
	collaboration = "TEXONO",
	title = "{Constraints on new physics with light mediators and generalized neutrino interactions via coherent elastic neutrino nucleus scattering}",
	eprint = "2502.20007",
	archivePrefix = "arXiv",
	primaryClass = "hep-ex",
	doi = "10.1103/63xf-t6fz",
	journal = "Phys. Rev. D",
	volume = "112",
	number = "3",
	pages = "035038",
	year = "2025"
}

@article{AristizabalSierra:2024nwf,
	author = "Aristizabal Sierra, D. and Mishra, N. and Strigari, L.",
	title = "{Implications of first neutrino-induced nuclear recoil measurements in direct detection experiments: Probing nonstandard interaction via CE{\ensuremath{\nu}}NS}",
	eprint = "2409.02003",
	archivePrefix = "arXiv",
	primaryClass = "hep-ph",
	doi = "10.1103/PhysRevD.111.055007",
	journal = "Phys. Rev. D",
	volume = "111",
	number = "5",
	pages = "055007",
	year = "2025"
}

@article{Coloma:2022umy,
	author = "Coloma, Pilar and Gonzalez-Garcia, M. C. and Maltoni, Michele and Pinheiro, Jo{\~a}o Paulo and Urrea, Salvador",
	title = "{Constraining new physics with Borexino Phase-II spectral data}",
	eprint = "2204.03011",
	archivePrefix = "arXiv",
	primaryClass = "hep-ph",
	reportNumber = "IFT-UAM/CSIC-22-14, IFIC/22-15, FTUV-22-0404.7998, YITP-SB-2022-05",
	doi = "10.1007/JHEP07(2022)138",
	journal = "JHEP",
	volume = "07",
	pages = "138",
	year = "2022",
	note = "[Erratum: JHEP 11, 138 (2022)]"
}

@article{Escrihuela:2023sfb,
	author = "Escrihuela, F. J. and Flores, L. J. and Miranda, O. G. and Rend{\'o}n, Javier and S{\'a}nchez-V{\'e}lez, R.",
	title = "{Examining the sensitivity of FASER to generalized neutrino interactions}",
	eprint = "2308.15630",
	archivePrefix = "arXiv",
	primaryClass = "hep-ph",
	doi = "10.1007/JHEP04(2024)102",
	journal = "JHEP",
	volume = "04",
	pages = "102",
	year = "2024"
}

@article{KATRIN:2024odq,
	author = "Aker, M. and others",
	collaboration = "KATRIN",
	title = "{First Constraints on General Neutrino Interactions Based on KATRIN Data}",
	eprint = "2410.13895",
	archivePrefix = "arXiv",
	primaryClass = "nucl-ex",
	doi = "10.1103/PhysRevLett.134.251801",
	journal = "Phys. Rev. Lett.",
	volume = "134",
	number = "25",
	pages = "251801",
	year = "2025"
}

@article{Denton:2024upc,
	author = "Denton, Peter B. and Giarnetti, Alessio and Meloni, Davide",
	title = "{Solar neutrinos and the strongest oscillation constraints on scalar NSI}",
	eprint = "2409.15411",
	archivePrefix = "arXiv",
	primaryClass = "hep-ph",
	doi = "10.1007/JHEP01(2025)097",
	journal = "JHEP",
	volume = "01",
	pages = "097",
	year = "2025"
}

@article{Liao:2025hcs,
	author = "Liao, Jiajun and Tang, Jian and Zhang, Bing-Long",
	title = "{Tensor interaction in coherent elastic neutrino-nucleus scattering}",
	eprint = "2502.10702",
	archivePrefix = "arXiv",
	primaryClass = "hep-ph",
	doi = "10.1103/19yb-tstx",
	journal = "Phys. Rev. D",
	volume = "112",
	number = "3",
	pages = "035036",
	year = "2025"
}

@article{DeRomeri:2025csu,
	author = "De Romeri, Valentina and Papoulias, Dimitrios K. and Sanchez Garcia, Gonzalo",
	title = "{Implications of the first CONUS+ measurement of coherent elastic neutrino-nucleus scattering}",
	eprint = "2501.17843",
	archivePrefix = "arXiv",
	primaryClass = "hep-ph",
	doi = "10.1103/PhysRevD.111.075025",
	journal = "Phys. Rev. D",
	volume = "111",
	number = "7",
	pages = "075025",
	year = "2025"
}

@article{Chattaraj:2025fvx,
	author = "Chattaraj, Ayan and Majumdar, Anirban and Srivastava, Rahul",
	title = "{Probing standard model and beyond with reactor CE{\ensuremath{\nu}}NS data of CONUS+ experiment}",
	eprint = "2501.12441",
	archivePrefix = "arXiv",
	primaryClass = "hep-ph",
	doi = "10.1016/j.physletb.2025.139438",
	journal = "Phys. Lett. B",
	volume = "864",
	pages = "139438",
	year = "2025"
}

@article{Alpizar-Venegas:2025wor,
	author = "Alp{\'\i}zar-Venegas, M. and Flores, L. J. and Peinado, Eduardo and V{\'a}zquez-J{\'a}uregui, E.",
	title = "{Exploring the standard model and beyond from the evidence of CE{\ensuremath{\nu}}NS with reactor antineutrinos in CONUS+}",
	eprint = "2501.10355",
	archivePrefix = "arXiv",
	primaryClass = "hep-ph",
	doi = "10.1103/PhysRevD.111.053001",
	journal = "Phys. Rev. D",
	volume = "111",
	number = "5",
	pages = "053001",
	year = "2025"
}

@article{Coloma:2023ixt,
	author = "Coloma, Pilar and Gonzalez-Garcia, M. C. and Maltoni, Michele and Pinheiro, Jo{\~a}o Paulo and Urrea, Salvador",
	title = "{Global constraints on non-standard neutrino interactions with quarks and electrons}",
	eprint = "2305.07698",
	archivePrefix = "arXiv",
	primaryClass = "hep-ph",
	reportNumber = "IFT-UAM/CSIC-23-47, IFIC/23-15, FTUV-23-0427.3710, YITP-SB-2023-05",
	doi = "10.1007/JHEP08(2023)032",
	journal = "JHEP",
	volume = "08",
	pages = "032",
	year = "2023"
}

@proceedings{Proceedings:2019qno,
    author = "Bhupal Dev, P. S. and others",
    title = "{Neutrino Non-Standard Interactions: A Status Report}",
    eprint = "1907.00991",
    archivePrefix = "arXiv",
    primaryClass = "hep-ph",
    reportNumber = "FERMILAB-CONF-19-299-T",
    doi = "10.21468/SciPostPhysProc.2.001",
    volume = "2",
    pages = "001",
    year = "2019"
}

@article{Flores:2021kzl,
    author = "Flores, L. J. and Nath, Newton and Peinado, Eduardo",
    title = "{CE\ensuremath{\nu}NS as a probe of flavored generalized neutrino interactions}",
    eprint = "2112.05103",
    archivePrefix = "arXiv",
    primaryClass = "hep-ph",
    doi = "10.1103/PhysRevD.105.055010",
    journal = "Phys. Rev. D",
    volume = "105",
    number = "5",
    pages = "055010",
    year = "2022"
}

@article{A:2022acy,
    author = "A., ShivaSankar K. and Majumdar, Anirban and Papoulias, Dimitrios K. and Prajapati, Hemant and Srivastava, Rahul",
    title = "{Implications of first LZ and XENONnT results: A comparative study of neutrino properties and light mediators}",
    eprint = "2208.06415",
    archivePrefix = "arXiv",
    primaryClass = "hep-ph",
    doi = "10.1016/j.physletb.2023.137742",
    journal = "Phys. Lett. B",
    volume = "839",
    pages = "137742",
    year = "2023"
}

@article{AtzoriCorona:2022jeb,
    author = "Atzori Corona, M. and Bonivento, W. M. and Cadeddu, M. and Cargioli, N. and Dordei, F.",
    title = "{New constraint on neutrino magnetic moment and neutrino millicharge from LUX-ZEPLIN dark matter search results}",
    eprint = "2207.05036",
    archivePrefix = "arXiv",
    primaryClass = "hep-ph",
    doi = "10.1103/PhysRevD.107.053001",
    journal = "Phys. Rev. D",
    volume = "107",
    number = "5",
    pages = "053001",
    year = "2023"
}

@article{Bahcall:1987jc,
    author = "Bahcall, John N. and Ulrich, Roger K.",
    title = "{Solar Models, Neutrino Experiments and Helioseismology}",
    reportNumber = "IASSNS-AST-87-1",
    doi = "10.1103/RevModPhys.60.297",
    journal = "Rev. Mod. Phys.",
    volume = "60",
    pages = "297--372",
    year = "1988"
}

@article{Baudis:2024jnk,
    author = "Baudis, Laura",
    title = "{DARWIN/XLZD: A future xenon observatory for dark matter and other rare interactions}",
    eprint = "2404.19524",
    archivePrefix = "arXiv",
    primaryClass = "astro-ph.IM",
    doi = "10.1016/j.nuclphysb.2024.116473",
    journal = "Nucl. Phys. B",
    volume = "1003",
    pages = "116473",
    year = "2024"
}

@Article{	  buchmuller:1986zs,
  author	= "Buchmuller, W. and Ruckl, R. and Wyler, D.",
  title		= "Leptoquarks in lepton quark collisions",
  journal	= "Phys. Lett.",
  volume	= "B191",
  year		= "1987",
  pages		= "442",
  slaccitation	= "%%CITATION = PHLTA,B191,442;%%"
}

@article{Crivellin:2019dwb,
	author = {Crivellin, Andreas and M\"uller, Dario and Saturnino, Francesco},
	title = "{Flavor Phenomenology of the Leptoquark Singlet-Triplet Model}",
	eprint = "1912.04224",
	archivePrefix = "arXiv",
	primaryClass = "hep-ph",
	reportNumber = "PSI-PR-19-25, ZU-TH 52/19",
	doi = "10.1007/JHEP06(2020)020",
	journal = "JHEP",
	volume = "06",
	pages = "020",
	year = "2020"
}

@article{DARWIN:2020bnc,
    author = "Aalbers, J. and others",
    collaboration = "DARWIN",
    title = "{Solar neutrino detection sensitivity in DARWIN via electron scattering}",
    eprint = "2006.03114",
    archivePrefix = "arXiv",
    primaryClass = "physics.ins-det",
    doi = "10.1140/epjc/s10052-020-08602-7",
    journal = "Eur. Phys. J. C",
    volume = "80",
    number = "12",
    pages = "1133",
    year = "2020"
}

@article{DeRomeri:2022twg,
	author = "De Romeri, V. and Miranda, O. G. and Papoulias, D. K. and Sanchez Garcia, G. and T\'ortola, M. and Valle, J. W. F.",
	title = "{Physics implications of a combined analysis of COHERENT CsI and LAr data}",
	eprint = "2211.11905",
	archivePrefix = "arXiv",
	primaryClass = "hep-ph",
	month = "11",
	year = "2022"
}

@article{Escrihuela:2021mud,
    author = "Escrihuela, F. J. and Flores, L. J. and Miranda, O. G. and Rend\'on, Javier",
    title = "{Global constraints on neutral-current generalized neutrino interactions}",
    eprint = "2105.06484",
    archivePrefix = "arXiv",
    primaryClass = "hep-ph",
    doi = "10.1007/JHEP07(2021)061",
    journal = "JHEP",
    volume = "07",
    pages = "061",
    year = "2021"
}

@Article{	  foot:1988aq,
  author	= "Foot, Robert and Lew, H. and He, X. G. and Joshi, Girish
		  C. ",
  title		= "{Seesaw neutrino masses induced by a triplet of leptons}",
  journal	= "Z. Phys.",
  volume	= "C44",
  year		= "1989",
  pages		= "441",
  xxxdoi	= "10.1007/BF01415558",
  slaccitation	= "%%CITATION = ZEPYA,C44,441;%%"
}

@article{Gargalionis:2020xvt,
	author = "Gargalionis, John and Volkas, Raymond R.",
	title = "{Exploding operators for Majorana neutrino masses and beyond}",
	eprint = "2009.13537",
	archivePrefix = "arXiv",
	primaryClass = "hep-ph",
	doi = "10.1007/JHEP01(2021)074",
	journal = "JHEP",
	volume = "01",
	pages = "074",
	year = "2021"
}

@misc{GranSasso_web,
  title		= "{Gran Sasso National Laboratory, INFN, Italy}",
  howpublished	= {http://www.lngs.infn.it/},
}

@article{Grimus:2006nb,
      author         = "Grimus, Walter",
      title          = "{Neutrino Physics - Models for Neutrino Masses and Lepton
                        Mixing}",
      booktitle      = "{Proceedings, School on Particle Physics, Gravity and
                        Cosmology}",
      journal        = "PoS",
      volume         = "P2GC",
      year           = "2006",
      pages          = "001",
      eprint         = "hep-ph/0612311",
      archivePrefix  = "arXiv",
      primaryClass   = "hep-ph",
      reportNumber   = "UWTHPH-2006-30",
      SLACcitation   = "%%CITATION = HEP-PH/0612311;%%"
}

@Article{	  hirsch:2004he,
  author	= "Hirsch, M. and Valle, J. W. F.",
  title		= "Supersymmetric origin of neutrino mass",
  journal	= "New J. Phys.",
  volume	= "6",
  year		= "2004",
  pages		= "76",
  eprint	= "hep-ph/0405015",
  slaccitation	= "%%CITATION = HEP-PH 0405015;%%"
}

@article{Kang:2010zza,
    author = "Kang, K. J. and Cheng, J. P. and Chen, Y. H. and Li, Y. J. and Shen, M. B. and Wu, S. Y. and Yue, Q.",
    editor = "Coccia, Eugenio and Pandola, Luciano and Fornengo, Nicolao and Aloisio, Roberto",
    title = "{Status and prospects of a deep underground laboratory in China}",
    doi = "10.1088/1742-6596/203/1/012028",
    journal = "J. Phys. Conf. Ser.",
    volume = "203",
    pages = "012028",
    year = "2010"
}

@article{LZ:2019sgr,
    author = "Akerib, D. S. and others",
    collaboration = "LZ",
    title = "{The LUX-ZEPLIN (LZ) Experiment}",
    eprint = "1910.09124",
    archivePrefix = "arXiv",
    primaryClass = "physics.ins-det",
    reportNumber = "FERMILAB-PUB-19-555-AE-E",
    doi = "10.1016/j.nima.2019.163047",
    journal = "Nucl. Instrum. Meth. A",
    volume = "953",
    pages = "163047",
    year = "2020"
}

@article{LZ:2022lsv,
    author = "Aalbers, J. and others",
    collaboration = "LZ",
    title = "{First Dark Matter Search Results from the LUX-ZEPLIN (LZ) Experiment}",
    eprint = "2207.03764",
    archivePrefix = "arXiv",
    primaryClass = "hep-ex",
    doi = "10.1103/PhysRevLett.131.041002",
    journal = "Phys. Rev. Lett.",
    volume = "131",
    number = "4",
    pages = "041002",
    year = "2023"
}

@article{LZ:2022ysc,
    author = "Aalbers, J. and others",
    collaboration = "LZ",
    title = "{Background determination for the LUX-ZEPLIN dark matter experiment}",
    eprint = "2211.17120",
    archivePrefix = "arXiv",
    primaryClass = "hep-ex",
    doi = "10.1103/PhysRevD.108.012010",
    journal = "Phys. Rev. D",
    volume = "108",
    number = "1",
    pages = "012010",
    year = "2023"
}

@article{LZ:2023poo,
    author = "Aalbers, J. and others",
    collaboration = "LZ",
    title = "{Search for new physics in low-energy electron recoils from the first LZ exposure}",
    eprint = "2307.15753",
    archivePrefix = "arXiv",
    primaryClass = "hep-ex",
    reportNumber = "FERMILAB-PUB-23-397-PPD",
    doi = "10.1103/PhysRevD.108.072006",
    journal = "Phys. Rev. D",
    volume = "108",
    number = "7",
    pages = "072006",
    year = "2023"
}

@article{Malinsky:2005bi,
      author         = "Malinsky, Michal and Romao, J.C. and Valle, J.W.F.",
      title          = "{Novel supersymmetric SO(10) seesaw mechanism}",
      journal        = "Phys.Rev.Lett.",
      volume         = "95",
      pages          = "161801",
      doi            = "10.1103/PhysRevLett.95.161801",
      year           = "2005",
      eprint         = "hep-ph/0506296",
      archivePrefix  = "arXiv",
      primaryClass   = "hep-ph",
      reportNumber   = "IFIC-05-28",
      SLACcitation   = "%%CITATION = HEP-PH/0506296;%%",
}

@article{Miranda:2015dra,
      author         = "Miranda, O. G. and Nunokawa, H.",
      title          = "{Non standard neutrino interactions: current status and
                        future prospects}",
      journal        = "New J. Phys.",
      volume         = "17",
      year           = "2015",
      number         = "9",
      pages          = "095002",
      doi            = "10.1088/1367-2630/17/9/095002",
      eprint         = "1505.06254",
      archivePrefix  = "arXiv",
      primaryClass   = "hep-ph",
      SLACcitation   = "%%CITATION = ARXIV:1505.06254;%%"
}

@article{Mohapatra1986bd,
  author	= "Mohapatra, R. N. and Valle, J. W. F.",
  title		= "Neutrino mass and baryon-number nonconservation in
		  superstring models",
  journal	= "Phys. Rev.",
  volume	= "D34",
  year		= "1986",
  pages		= "1642",
  slaccitation	= "%%CITATION = PHRVA,D34,1642;%%"
}

@article{Mount:2017qzi,
    author = "Mount, B. J. and others",
    title = "{LUX-ZEPLIN (LZ) Technical Design Report}",
    eprint = "1703.09144",
    archivePrefix = "arXiv",
    primaryClass = "physics.ins-det",
    reportNumber = "LBNL-1007256, FERMILAB-TM-2653-AE-E-PPD",
    month = "3",
    year = "2017"
}

@article{Ohlsson:2012kf,
      author         = "Ohlsson, Tommy",
      title          = "{Status of non-standard neutrino interactions}",
      journal        = "Rept. Prog. Phys.",
      volume         = "76",
      year           = "2013",
      pages          = "044201",
      doi            = "10.1088/0034-4885/76/4/044201",
      eprint         = "1209.2710",
      archivePrefix  = "arXiv",
      primaryClass   = "hep-ph",
      SLACcitation   = "%%CITATION = ARXIV:1209.2710;%%"
}

@article{PandaX:2022ood,
    author = "Zhang, Dan and others",
    collaboration = "PandaX",
    title = "{Search for Light Fermionic Dark Matter Absorption on Electrons in PandaX-4T}",
    eprint = "2206.02339",
    archivePrefix = "arXiv",
    primaryClass = "hep-ex",
    doi = "10.1103/PhysRevLett.129.161804",
    journal = "Phys. Rev. Lett.",
    volume = "129",
    number = "16",
    pages = "161804",
    year = "2022"
}

@article{PandaX-4T:2021bab,
    author = "Meng, Yue and others",
    collaboration = "PandaX-4T",
    title = "{Dark Matter Search Results from the PandaX-4T Commissioning Run}",
    eprint = "2107.13438",
    archivePrefix = "arXiv",
    primaryClass = "hep-ex",
    doi = "10.1103/PhysRevLett.127.261802",
    journal = "Phys. Rev. Lett.",
    volume = "127",
    number = "26",
    pages = "261802",
    year = "2021"
}

@article{Ruppin:2014bra,
    author = "Ruppin, F. and Billard, J. and Figueroa-Feliciano, E. and Strigari, L.",
    title = "{Complementarity of dark matter detectors in light of the neutrino background}",
    eprint = "1408.3581",
    archivePrefix = "arXiv",
    primaryClass = "hep-ph",
    doi = "10.1103/PhysRevD.90.083510",
    journal = "Phys. Rev. D",
    volume = "90",
    number = "8",
    pages = "083510",
    year = "2014"
}

@article{Schechter1980gr,
      author         = "Schechter, J. and Valle, J.W.F.",
      title          = "{Neutrino Masses in SU(2) x U(1) Theories}",
      journal        = "Phys.Rev.",
      volume         = "D22",
      pages          = "2227",
      doi            = "10.1103/PhysRevD.22.2227",
      year           = "1980",
      reportNumber   = "SU-4217-167, COO-3533-167",
      SLACcitation   = "%%CITATION = PHRVA,D22,2227;%%",
}

@article{Villante:2020adi,
    author = "Villante, Francesco L. and Serenelli, Aldo",
    title = "{The relevance of nuclear reactions for Standard Solar Models construction}",
    eprint = "2101.03077",
    archivePrefix = "arXiv",
    primaryClass = "astro-ph.SR",
    doi = "10.3389/fspas.2020.618356",
    journal = "Front. Astron. Space Sci.",
    volume = "7",
    pages = "112",
    year = "2021"
}

@article{XENON:2010xwm,
    author = "Aprile, E. and others",
    collaboration = "XENON",
    title = "{Design and Performance of the XENON10 Dark Matter Experiment}",
    eprint = "1001.2834",
    archivePrefix = "arXiv",
    primaryClass = "astro-ph.IM",
    doi = "10.1016/j.astropartphys.2011.01.006",
    journal = "Astropart. Phys.",
    volume = "34",
    pages = "679--698",
    year = "2011"
}

@article{XENON:2020rca,
    author = "Aprile, E. and others",
    collaboration = "XENON",
    title = "{Excess electronic recoil events in XENON1T}",
    eprint = "2006.09721",
    archivePrefix = "arXiv",
    primaryClass = "hep-ex",
    doi = "10.1103/PhysRevD.102.072004",
    journal = "Phys. Rev. D",
    volume = "102",
    number = "7",
    pages = "072004",
    year = "2020"
}

@article{XENON:2022ltv,
    author = "Aprile, E. and others",
    collaboration = "XENON",
    title = "{Search for New Physics in Electronic Recoil Data from XENONnT}",
    eprint = "2207.11330",
    archivePrefix = "arXiv",
    primaryClass = "hep-ex",
    doi = "10.1103/PhysRevLett.129.161805",
    journal = "Phys. Rev. Lett.",
    volume = "129",
    number = "16",
    pages = "161805",
    year = "2022"
}

@article{XLZD:2024nsu,
    author = "Aalbers, J. and others",
    collaboration = "XLZD",
    title = "{The XLZD Design Book: towards the next-generation liquid xenon observatory for dark matter and neutrino physics}",
    eprint = "2410.17137",
    archivePrefix = "arXiv",
    primaryClass = "hep-ex",
    doi = "10.1140/epjc/s10052-025-14810-w",
    journal = "Eur. Phys. J. C",
    volume = "85",
    number = "10",
    pages = "1192",
    year = "2025"
}

@article{Zhao:2020ezy,
    author = "Zhao, Li and Liu, Jianglai",
    title = "{Experimental search for dark matter in China}",
    eprint = "2004.04547",
    archivePrefix = "arXiv",
    primaryClass = "astro-ph.IM",
    doi = "10.1007/s11467-020-0960-x",
    journal = "Front. Phys. (Beijing)",
    volume = "15",
    number = "4",
    pages = "44301",
    year = "2020"
}

@article{Khan:2019jvr,
    author = "Khan, Amir N. and Rodejohann, Werner and Xu, Xun-Jie",
    title = "{Borexino and general neutrino interactions}",
    eprint = "1906.12102",
    archivePrefix = "arXiv",
    primaryClass = "hep-ph",
    reportNumber = "FERMILAB-PUB-19-348-T",
    doi = "10.1103/PhysRevD.101.055047",
    journal = "Phys. Rev. D",
    volume = "101",
    number = "5",
    pages = "055047",
    year = "2020"
}

@article{Rodejohann:2017vup,
    author = "Rodejohann, Werner and Xu, Xun-Jie and Yaguna, Carlos E.",
    title = "{Distinguishing between Dirac and Majorana neutrinos in the presence of general interactions}",
    eprint = "1702.05721",
    archivePrefix = "arXiv",
    primaryClass = "hep-ph",
    doi = "10.1007/JHEP05(2017)024",
    journal = "JHEP",
    volume = "05",
    pages = "024",
    year = "2017"
}

@article{AristizabalSierra:2018eqm,
    author = "Aristizabal Sierra, D. and De Romeri, Valentina and Rojas, N.",
    title = "{COHERENT analysis of neutrino generalized interactions}",
    eprint = "1806.07424",
    archivePrefix = "arXiv",
    primaryClass = "hep-ph",
    doi = "10.1103/PhysRevD.98.075018",
    journal = "Phys. Rev. D",
    volume = "98",
    pages = "075018",
    year = "2018"
}

@article{Bischer:2019ttk,
    author = "Bischer, Ingolf and Rodejohann, Werner",
    title = "{General neutrino interactions from an effective field theory perspective}",
    eprint = "1905.08699",
    archivePrefix = "arXiv",
    primaryClass = "hep-ph",
    doi = "10.1016/j.nuclphysb.2019.114746",
    journal = "Nucl. Phys. B",
    volume = "947",
    pages = "114746",
    year = "2019"
}

@article{Bischer:2018zcz,
    author = "Bischer, Ingolf and Rodejohann, Werner",
    title = "{General Neutrino Interactions at the DUNE Near Detector}",
    eprint = "1810.02220",
    archivePrefix = "arXiv",
    primaryClass = "hep-ph",
    doi = "10.1103/PhysRevD.99.036006",
    journal = "Phys. Rev. D",
    volume = "99",
    number = "3",
    pages = "036006",
    year = "2019"
}

@article{Han:2020pff,
    author = "Han, Tao and Liao, Jiajun and Liu, Hongkai and Marfatia, Danny",
    title = "{Scalar and tensor neutrino interactions}",
    eprint = "2004.13869",
    archivePrefix = "arXiv",
    primaryClass = "hep-ph",
    reportNumber = "PITT-PACC-2003",
    doi = "10.1007/JHEP07(2020)207",
    journal = "JHEP",
    volume = "07",
    pages = "207",
    year = "2020"
}

@article{Farzan:2017xzy,
    author = "Farzan, Y. and Tortola, M.",
    title = "{Neutrino oscillations and Non-Standard Interactions}",
    eprint = "1710.09360",
    archivePrefix = "arXiv",
    primaryClass = "hep-ph",
    doi = "10.3389/fphy.2018.00010",
    journal = "Front. in Phys.",
    volume = "6",
    pages = "10",
    year = "2018"
}

@article{Bahcall:1994cf,
    author = "Bahcall, John N.",
    title = "{The Be-7 solar neutrino line: A Reflection of the central temperature distribution of the sun}",
    eprint = "astro-ph/9401024",
    archivePrefix = "arXiv",
    reportNumber = "IASSNS-AST-93-40",
    doi = "10.1103/PhysRevD.49.3923",
    journal = "Phys. Rev. D",
    volume = "49",
    pages = "3923--3945",
    year = "1994"
}

@article{Bahcall:1996qv,
    author = "Bahcall, John N. and Lisi, E. and Alburger, D. E. and De Braeckeleer, L. and Freedman, S. J. and Napolitano, J.",
    title = "{Standard neutrino spectrum from B-8 decay}",
    eprint = "nucl-th/9601044",
    archivePrefix = "arXiv",
    reportNumber = "IASSNS-AST-96-28, IASSNS-AST-9547",
    doi = "10.1103/PhysRevC.54.411",
    journal = "Phys. Rev. C",
    volume = "54",
    pages = "411--422",
    year = "1996"
}

@article{ParticleDataGroup:2024cfk,
    author = "Navas, S. and others",
    collaboration = "Particle Data Group",
    title = "{Review of particle physics}",
    doi = "10.1103/PhysRevD.110.030001",
    journal = "Phys. Rev. D",
    volume = "110",
    number = "3",
    pages = "030001",
    year = "2024"
}

@article{Chen:2016eab,
    author = "Chen, Jiunn-Wei and Chi, Hsin-Chang and Liu, C. -P. and Wu, Chih-Pan",
    title = "{Low-energy electronic recoil in xenon detectors by solar neutrinos}",
    eprint = "1610.04177",
    archivePrefix = "arXiv",
    primaryClass = "hep-ex",
    reportNumber = "NCTS-ECP-1601, MIT-CTP-4843",
    doi = "10.1016/j.physletb.2017.10.029",
    journal = "Phys. Lett. B",
    volume = "774",
    pages = "656--661",
    year = "2017"
}

\end{document}